\documentclass[12pt]{iopart}
\usepackage{iopams}
\usepackage{bm}% bold math
\usepackage{hyperref} % add hypertext capabilities
\hypersetup{colorlinks = true, linkcolor = blue, anchorcolor = blue, citecolor = blue, filecolor = blue, urlcolor = blue}

\begin{document}

\title{Time evolution of spin singlet in static homogeneous exchange and magnetic fields}
\footnote{\href{http://fnte.ilt.kharkov.ua/join.php?fn=/fnt/pdf/47/47-10/f47-0895e.pdf}{\textit{Fiz. Nizk. Temp.} \textbf{47}, 895–902 (2021)};  \href{https://doi.org/10.1063/10.0006061}{\textit{Low Temp. Phys.} \textbf{47}, 823 (2021)}.}
\author{S.~V.~Kuplevakhsky, S.~V.~Bengus}
\address{B.~Verkin Institute for Low Temperature Physics and Engineering of the National Academy of Sciences of Ukraine, 47 Nauky Avenue, Kharkiv 61103, Ukraine}
\ead{sbengus@ilt.kharkov.ua}

%\date{March 15, 2021}% It is always \today, today, but any date may be explicitly specified

\begin{abstract}
Within the framework of an idealized theoretical model, we study the effect of external static homogeneous exchange and magnetic field on the spin part of the singlet wave function of two electrons. We begin by revising the traditional (textbook) approach to the spin singlet. Basing our own approach solely on the property of invariance under rotations of the coordinate system and using the theory of spinor invariants, we derive a generalized representation of the spin singlet whose main feature is that the spins are in mutually time-reversed states. We show that exactly this feature predetermines the actual form of the Hamiltonian of interaction with the external field and stipulates time evolution of the singlet. Some applications of these results to the theory of superconductivity and spin chemistry are presented. In particular, it is shown that the case of ferromagnetic superconductors constitutes a good illustration of the validity of our quantum-mechanical consideration.\\ \\
\noindent \textit{Keywords}: spin singlet, time reversal, evolution, exchange and magnetic fields.
\end{abstract}
%===============
\pacs{03.65.-w, 74.20.Fg, 82.20.Ej}
%Classification Scheme.
% 03.65.-w Quantum mechanics
% 74.20.Fg BCS theory and its development
% 82.20.Ej Quantum theory of reaction cross section
%===============
%\submitto{Thtydt}

\maketitle

\section{Introduction}

The aim of this paper is to study theoretically the effect of external static homogeneous exchange and magnetic fields on the spin part of a singlet wave function of two electrons. For fear of possible misreading, we shall first of all formulate our exact statement of the problem. The idealized model accepted in this paper does not take into account any electromagnetic or exchange interactions between the electrons of the singlet. Moreover, to make our consideration uniform, we completely disregard orbital degrees of freedom and concentrate only on spin dynamics. Surprisingly, although the thus stated problem concerns the basics of quantum mechanics and has important applications in related sciences, it is not discussed in standard textbooks \cite{Schiff,DBohm,Flugge,Landau,ABohm,Sudbery}, and we are unaware of the correct solution to it in literature. For example, if we choose the spin quantization axis to be perpendicular to the external field, we notice that the probabilities of definite spin orientations oscillate with time and that spin flips occur \cite{Flugge}. These intuitive conjectures about the behavior of the spin singlet in the presence of external fields will be verified, refined on and developed by means of rigorous mathematical methods in the sections of the paper that follow. 

In particular, we begin \textbf{Section 2} with an analysis of certain drawbacks of
the traditional (textbook) \cite{Schiff,DBohm,Flugge,Landau,ABohm,Sudbery}
representation of the spin singlet. After that, based on the theory of spinor invariants
\cite{Cartan,Brinkman}, we derive a generalized representation of the spin singlet
which is free from the drawbacks of the traditional one: the generalized representation is
explicitly invariant under rotations of the coordinate system. The main
feature of the generalized representation is that the spins are in mutually
time-reversed states. Relationship to the representation of the spin singlet
as a normalized metric spinor is established.

In \textbf{Section 3}, we use the results of section \textbf{Section 2} to study the evolution of the
spin singlet. An exact time dependent spin wave function is derived. This
wave function exhibits periodic conversions from the spin singlet to the
zero component of the spin triplet along the external field. Periodic
permutations of the spins of the singlet, caused by spin flips, are also
envisaged.

In \textbf{Section 4}, we consider the application of the results of the previous
section to the theory of ferromagnetic superconductors and spin chemistry.
Some mathematical details related to the results of sections \textbf{Sections 2-4} are
relegated to \textbf{Appendices\,A} and \textbf{B}. In section \textbf{Section 5}, key results of the paper are
discussed and several conclusions are drawn.

%=>Section2
\section{The generalized representation of the spin singlet}

The correlation between the spins of the singlet clearly manifests itself in the property of invariance under rotations of the coordinate system. To explain the situation, we begin by drawing the reader's attention to some little-known mathematical aspects of the singlet wave function, not mentioned in standard textbooks (see, e. g., [1-6]). 

Traditionally, the singlet wave function is written down as an antisymmetric
linear combination of the eigenfunctions of one of the Cartesian components
of the total spin $S=s_{1}+s_{2}$, corresponding to a zero eigenvalue of the operator ${S^2}$:%
\begin{equation}
\Psi _{S}\left( 1,2\right) =\frac{1}{\sqrt{2}}\left( \Psi _{\alpha+}\otimes
 \Psi _{\alpha-}-\Psi _{\alpha-}\otimes \Psi _{\alpha+}\right). \label{eq:1}
\end{equation}
Here, $\alpha=x;y;z$; the sign $\otimes $ denotes a direct product of two-dimensional
Hilbert spaces of spin 1 (on the left) and spin 2 (on the right); $\Psi _{\alpha+}
$ and $\Psi _{\alpha-}$ are the eigenfunctions of the corresponding Pauli matrices.

Using the properties of the time-reversal operator \textit{K} \cite{Messiah}, we may obtain
the following generalized representation of the singlet state:
\begin{eqnarray}
\Psi _{S}\left( 1,2\right) 
=\frac{1}{\sqrt{2}}
\left( K\Psi _{\alpha-}\otimes \Psi _{\alpha-}+
K\Psi _{\alpha+} \otimes \Psi _{\alpha+} \right),  \label{eq:2}
\end{eqnarray}
where $K\Psi _{\mathbf{\alpha}-}=\Psi _{\mathbf{\alpha}+}$ and 
$K\Psi _{\mathbf{\alpha}+}=-\Psi _{\mathbf{\alpha}-}$. 

We want to say that the Eq.(\ref{eq:2}) are not merely a new representation of the singlet state, different from the traditional one.
It emphasizes only that the singlet state is formed by two spin states that are mutually reversed in time.

%SECTION 3
\section{Time evoution of the spin singlet}
Now we are fully prepared to return to our main problem: the evolution
of the singlet state. If the spins were independent, the dynamics of both of
them would be generated by the same single-particle Hamiltonian
    %3=>
\begin{equation}
\mathcal{H}=-\sigma _{z}J. \label{eq:3}
\end{equation}%
Note that in the case of an exchange field, which is parallel to $z$-direction, $J$ is just its value; in the case of a magnetic field
$H$, $\displaystyle J= -g\mu _{b}H_{z}$ with
$\displaystyle \mu_{b}$
is the Bohr magneton.
The evolution operator for an initial state $\Psi $ is
    %4=>
\begin{equation}
U\left( t\right) =\exp \left( -i\frac{\mathcal{H}}{\hbar}t\right) ,  \label{eq:4}
\end{equation}%
The evolution operator for the time-revesred state $K \Psi$ is \cite{Messiah}
    %5=>
\begin{equation}
U _{\textnormal{rev}}\left( t\right) =KU\left( -t\right) K^{+}=\exp \left( -i\frac{K\mathcal{H}K^{+}}{\hbar}t\right) .  \label{eq:5}
\end{equation}
Thus, the evolution operator for the spin singlet (2) has the following form:
    %6=>
\begin{equation}
U_{1,2} \left( t \right) =\exp \left( -i\frac{K\mathcal{H}K^{+}}{\hbar}t \right) \otimes
\exp \left( -i \frac{\mathcal{H}}{\hbar} t \right), \label{eq:6}
\end{equation}
For the perpendicular magnetic field the corresponding time-dependent two-spin state has the form
    %7=>
\begin{equation}
\Psi(1,2;t) 
= a(t) \Psi_S(1,2) + b(t)\Psi_{T}(1,2)\nonumber \label{eq:7}
\end{equation}%
where $\Psi_T (1,2)$ is the triplet two-spin function with $z$-projection of total spin $S^Z=0$
    %8=>
\begin{equation}
\Psi_{T}(1,2) =\frac{1}{\sqrt{2}}(\Psi _{\mathbf{\alpha}+}\otimes\Psi _{\mathbf{\alpha}-}+\Psi _{\mathbf{\alpha}-}\otimes\Psi _{\mathbf{\alpha}+}) \nonumber \label{eq:8}
\end{equation}
Among other things, relation (\ref{eq:6}) imply that the
actual interaction Hamiltonian for the spin singlet is not
    %9=>
\begin{equation}
\mathcal{H}\otimes I+I\otimes \mathcal{H},  \label{eq:9}
\end{equation}%
as would be the case for two independent spins, but rather
    %10=>
\begin{equation}
\mathcal{H}\otimes I+I\otimes K\mathcal{H}K^{+}  \label{eq:10}
\end{equation}%
or
    %11=>
\begin{equation}
K\mathcal{H}K^{+}\otimes I+\mathcal{H}\otimes I,  \label{eq:11}
\end{equation}%
where $I$ is a unit operator. Relations (\ref{eq:10}) and (\ref{eq:11}) take explicitly into account the
correlation between the spins of the singlet.

Consider first the representation (\ref{eq:2}) and the
evolution operator (\ref{eq:6}). Although the explicit form of the time-dependent state
    %12=> 
\begin{eqnarray}
\Psi \left( 1,2;t\right)&=&U_{1,2}\left( t\right) \Psi _{S}\left( 1,2\right) \nonumber \\
&\equiv& \frac{1}{\sqrt{2}} 
\left[ U\left( t\right) \Psi _{\mathbf{\hat{n}}+}\otimes U_{\textnormal{rev}}\left( t\right)K\Psi _{\mathbf{\hat{n}}+} \right. \nonumber \\
&& \,\,\,\,\,+\,\left. U\left( t\right)\Psi _\mathbf{\hat{n}-}\otimes U_{\textnormal{rev}}\left( t \right )K\Psi _{\mathbf{\hat{n}}-} \right]   \label{eq:12}
\end{eqnarray}
can be evaluated for an arbitrary direction of the vector $\mathbf{\hat{n}}$ in Eq.(\ref{eq:12}), from the
point of view of physical interpretation, it is reasonable to take $\mathbf{\hat{n}}$ 
perpendicular to the direction of the field: say, $\mathbf{\hat{n}=\hat{x}}$.
In this way, we immediately arrive at the following set of expressions:
     %13=>
\begin{eqnarray}
\Psi \left( 1,\!2;t\right)  \!=\!\! \left\{\!
\begin{array}{l}
\!a\!\left( t\right) \Psi _{S}\!\left( 1,2\right) +b\left( t\right) \Psi_{T,S_{z}=0}\!\left( 1,2\right)\! , a\left( t\right) \!>0; \\ 
\!\left\vert a\!\left( t\right)\right\vert  \Psi _{S}\!\left( 2,\!1\right) +b\left(t\right) \Psi _{T,S_{z}=0}\!\left( 2,\!1\right)\!, a\!\left( t\right) \!<0;
\end{array}%
\right. \label{eq:13}
\end{eqnarray}
     %14=>
\begin{eqnarray}
&&\Psi  _{S} \left( 1,2 \right) = -\Psi  _{S} \left( 2,1 \right) \nonumber \\
&&= -\frac{1}{2}\left[ \left(
\begin{array}{c}
1 \\ 
1
\end{array}%
\right) \otimes \left( 
\begin{array}{c}
1 \\ 
-1
\end{array}%
\right) -\left( 
\begin{array}{c}
1 \\ 
-1
\end{array}%
\right) \otimes \left( 
\begin{array}{c}
1 \\ 
1
\end{array}%
\right) \right] \nonumber \\
&&= \frac{1}{2}\left[ \left(
\begin{array}{c}
1 \\ 
0
\end{array}%
\right) \otimes \left( 
\begin{array}{c}
0 \\ 
1
\end{array}%
\right) -\left( 
\begin{array}{c}
0 \\ 
1
\end{array}%
\right) \otimes \left( 
\begin{array}{c}
1 \\ 
0
\end{array}%
\right) \right], \label{eq:14} \\
    %15=>
&&\Psi _{T,S_z =0} \left( 1,2 \right) = \Psi _{T,S_z =0} \left( 2,1 \right) \nonumber \\
&&=\frac{1}{2} \left[ \left( 
\begin{array}{c}
     1 \\ 
     0
\end{array}%
\right) \otimes \left( 
\begin{array}{c}
     0 \\ 
     1
\end{array}%
\right) +\left( 
\begin{array}{c}
     0 \\ 
     1
\end{array}%
\right) \otimes \left( 
\begin{array}{c}
     1 \\ 
     0
\end{array}%
\right)  \right]; \label{eq:15}
\end{eqnarray}
\begin{eqnarray}
    %16=>
a \left( t\right)~=~&\cos& \left( \frac{2Jt}{\hbar} \right) 
  \equiv 2
  \left[ \frac{1}{2}-\sin ^{2}\left( \frac{Jt}{\hbar}\right) \right] , \label{eq:16} \\
    %17=>
b \left( t\right)~=~&i\sin& \left( \frac{2Jt}{\hbar} \right)
  \equiv
   i2\sin \left( \frac{Jt}{\hbar}\right)\cos \left( \frac{Jt}{\hbar} \right). \label{eq:17}
\end{eqnarray}%
Here, $\Psi _{T,S_z =0}\left( 1,2\right) $ is the component of the triplet state corresponding to 
$S_{z}=0$; $a\left( t\right) $ and $b\left( t\right) $ are
the probability amplitudes of the states $\Psi _{S}$ and $\Psi _{T,0}$,
respectively; $\displaystyle \sin ^{2}\left( \frac{Jt}{\hbar}\right)$
is the probability of a spin flip in a perpendicular field $J$ \cite{Flugge};
$\displaystyle \sin \left( \frac{Jt}{\hbar} \right)$ and $\displaystyle \cos \left( \frac{Jt}{\hbar} \right)$
are the probability amplitudes of a spin flip and of the absense of a spin flip, respectively.
Furthermore, $\displaystyle\left\vert \frac{1}{2}-\sin^{2}\left( \frac{Jt}{\hbar}\right) \right\vert $
is the probability of a definite spin orientation.

In addition, we want to emphasize that time
dependence of the probability amplitude $a=a\left( t\right) $ reflects the dynamics of the time-reversal operator
$K=K\left( t\right) $, which is clear from the representations derived in \textbf{Appendix\,A}:%
   %18=>
\begin{equation}
a\left( t\right) \!=\!\frac{1}{2}\mathop{\mathrm{Sp}}\left[ K\left( t\right) K^{+}\!\left( 0\right) \right]\!
=\!\frac{1}{2}\mathop{\mathrm{Sp}}\left[ K\left( -t\right) K^{+}\!\left( 0\right) \right] ;  \label{eq:18}
\end{equation}
   %19=>
\begin{equation}
K\left( t\right)  =\exp \left( i\frac{\mathcal{H}}{\hbar}t\right) K\left( 0\right) \exp \left( i\frac{\mathcal{H}}{\hbar}t\right) .  \label{eq:19}
\end{equation}
Here, the time-reversal operators are, of course, written down in the Heisenberg representation. Equal sign in the arguments of both the exponents in Eq.(\ref{eq:19}) is due to the antilinearity of $K$: $Ki=-iK$.

As can be seen from Eqs.(\ref{eq:13})-(\ref{eq:17}), when an external non-time-reversible field
is "switched on" at $t=0$, the initial singlet state $\Psi _{S}$ starts to
decay gradually, whereas the zero component of the triplet state along the external field,
$\Psi _{T,S_z =0}$, is emerging owing to spin flips induced by the field.
At $\displaystyle t=\frac{\pi \hbar}{4J}$ a permutation of the spins 1 and 2 occurs, which is reflected in the second line 
of Eq.(\ref{eq:13}). At $\displaystyle t=\frac{3\pi \hbar}{4J}$ a new permutation of the spins occurs. 
Formally, the process is periodic with the period $\displaystyle T=\frac{\pi \hbar}{J}$.

Certainly, in view of idealized character of our model (see the \textbf{Introduction}) 
the possibility of the observation of the above-described
quantum-mechanical effects in real electron systems strongly depends on
concrete physical situations. For example, periodic conversions 
$\Psi_{S}\rightarrow \Psi _{T,S_{z}=0}$, envisaged by Eqs. (\ref{eq:13})-(\ref{eq:17}),
are prohibited in homogeneous ferromagnetic superconductors \textbf{Section 4.1}).
By contrast, such conversions are experimentally observed in some situations encountered in
spin chemistry (\textbf{Section 4.2}).

If we now take the representation of (\ref{eq:2}) and the
evolution operator (\ref{eq:6}), the result for the corresponding
time-dependent state will be straightforward:
   %20=>
\begin{eqnarray}
&\Psi ^{\ast }\left( 1,2;t\right) \equiv U_{1,2}^{*}\left( t\right)\Psi _{S}\left( 1,2\right) =K_{0}U_{1,2}\left( t\right) \Psi _{S}\left(1,2\right)  \nonumber \\
&\!\!=\!\left\{ 
\begin{array}{c}
\!\!a\left( t\right) \Psi _{S}\left( 1,2\right) +b^{*}\!\left( t\right) \Psi_{T,S_{z}=0}\!\left( 1,2\right)\!, a\left( t\right) >0; \\ 
\!\!\left\vert a\left( t\right) \right\vert \Psi _{S}\left( 2,1\right)+b^{*}\!\left( t\right) \Psi _{T,S_{z}=0}\!\left( 2,1\right)\!, a\left(t\right) <0.%
\end{array}%
\right.  \label{eq:20}
\end{eqnarray}%

Given that the function $\Psi ^{*}\left( 1,2;t\right)$ is the complex conjugate of the function $\Psi \left( 1,2;t\right)$
and differs from the latter only by complex conjugation of the probability amplitude 
$b\left( t\right)$ [$\displaystyle b^{*}\left( t\right) =-i\sin \left( \frac{2Jt}{\hbar}\right)$],
both $\Psi \left( 1,2;t\right) $ and $\Psi ^{*}\left( 1,2;t\right) $ describe the same physical situation, as could be expected.

%SECTION 4
\section{Applications}
The quantum-mechanical results of the two previous sections have immediate
applications in related sciences, namely the theory of superconductivity and
spin chemistry. We begin with the theory of superconductivity.

%SECTION 4.1
\subsection{Coexistence of superconductivity and ferromagnetism}
As was first observed by Anderson a long time ago \cite{Anderson59} 
(see also Ref.\cite{Anderson84}), in the BCS
theory of superconductivity \cite{Bardeen}, superconducting correlations (or
Cooper pairs) are formed by electron states that are mutually reversed in time,
e.g., $\left\vert \mathbf{p\uparrow }\right\rangle $ and $\left\vert -
\mathbf{p\downarrow }\right\rangle $ if the electron momentum $\mathbf{p}$
is a good quantum number. Unfortunately, it seems that implications of this
observation for ferromagnetic superconductors have not been understood in
the literature. As an explanation, we consider the linearized equation for
the superconducting order parameter $\Delta =\Delta \left( \mathbf{r}\right) $, 
valid near the transition curve between the superconducting and normal phases 
$T_{c}=T_{c}\left( J\right)$ (provided the transition is of second order):
   %21=>
\begin{eqnarray}
\Delta \left( \mathbf{r}\right) &=&\int d\mathbf{r}^{\prime 3}\mathcal{K}\left( \mathbf{r,r}^{\prime }\right) 
\Delta \left( \mathbf{r}^{\prime }\right) , \label{eq:21}\\
\mathcal{K}\left( \mathbf{r,r}^{\prime }\right) &=&\int \frac{d\mathbf{p}^{3}}{\left(2\pi \hbar \right) ^{3}}
\exp \left[  \frac{i\mathbf{p}}{\hbar}\left( \mathbf{r}-\mathbf{r}^{\prime }\right) \right] \mathcal{K}\left( p\right) .  \nonumber
\end{eqnarray}
As regards some details, see e.g., the old reviews \cite{Saint-James,Izyumov} and references therein.

A Fourier transform of Eq.(\ref{eq:21}) to the momentum space was employed in the
literature \cite{Saint-James,Izyumov} to evaluate peculiar behavior of the second-order transition curve 
that had a branching point designating the
origin of a first-order phase transition, but we will not discuss this issue
here. Neither will we ponder on the problem of existence or non-existence
of the so-called FFLO phase (see the original papers \cite{Fulde,Larkin} and the
review \cite{Eschrig}): this problem is also beyond the scope of our paper.
Instead, we will focus on those mathematical properties of the integral
kernel $\mathcal{K}\left( \mathbf{r,r}^{\prime }\right)$ that are intimately connected
with our quantum-mechanical results and not reflected in the existing
literature.

The quantity $\mathcal{K}\left( p\right)$
is given in Ref.\cite{Saint-James} in the quasi-classical approximation
(when $\displaystyle \max \left\{ \frac{T_{c}}{E_{F}},\frac{J}{E_{F}}\right\} \ll 1$, with 
$E_{F}$ being the Fermi energy) for the two extreme cases, namely: the
``clean'' limit (no impurities) and the ``dirty'' limit (a chaotic distribution
of non-magnetic impurities). However, to elucidate the effect of the
exchange field, we have to resort to the coordinate representation of the
integral kernel that describes the propagation of superconducting
correlations between the points $\mathbf{r}^{\prime }$ and $\mathbf{r}$.
Thus, in the ``clean'' limit we have:
   %22=>
\begin{eqnarray}
\mathcal{K}\left( p\right)  &=&\frac{2\pi N\left( 0\right) \left\vert g \right\vert T_{c}}{pv_{F}} \sum\limits_{\omega _{n}>0} 
\left[ \arctan{\left( \frac{pv_{F}-2J}{2\omega _{n}} \right)} \right. \nonumber \\
&+&\left. \arctan{\left( \frac{pv_{F}+2J}{2\omega _{n}} \right)}\right] ,  \label{eq:22}
\end{eqnarray}
   %23=>
\begin{eqnarray}
\mathcal{K}\left( \mathbf{r,r}^{\prime }\right) &=& \frac{N\left( 0\right) \left\vert g \right\vert T_{c}}{\hbar v_{F}
\left\vert \mathbf{r} - \mathbf{r}^{\prime }\right\vert ^{2}}
\sum\limits_{\omega _{n}>0} \left[ 1-2\sin ^{2}\left( \frac{\left\vert \mathbf{r}-\mathbf{r}^{\prime}\right\vert J}{\hbar v_{F}}\right)\right] \nonumber \\
&\times& \exp \left( -\frac{2\omega _{n}\left\vert \mathbf{r}-\mathbf{r}^{\prime}\right\vert}{\hbar v_{F}}\right), \label{eq:23}
\end{eqnarray}
where $\displaystyle N\left( 0\right)=\frac{mp_{F}}{2\pi^{2}\hbar^{3}}$ is the density of states at the Fermi level in the normal phase, 
$v_{F} $ is the Fermi velocity, $\left\vert g\right\vert $ is the value of the
constant of effective electron-electron interaction, and $\omega _{n}=\left( 2n+1\right) \pi T_{c}$
($n=0,\pm 1,\pm 2,...$). As can be easily seen, the preexponential factor in
the square brackets in Eq.(\ref{eq:23}) is nothing but an image (in a rigorous mathematical sense) of the
probability amplitude $a \left( t\right)$ [Eq.(\ref{eq:18})]. Indeed, 
physically, the quasi-classical approximation implies that each electron of
the Cooper pair is represented by a wave packet \cite{DBohm,Landau}
formed by the states with the momenta 
$\displaystyle p \;{\in} \left( 
p_{F}-\max 
\left\{ \frac{T_{c}}{v_{F}},\frac{J}{v_{F}}\right\} \!, \; 
p_{F}+\max 
\left\{ \frac{T_{c}}{v_{F}},\frac{J}{v_{F}}\right\} \right) $.
The centres of these packets move at the velocity $v_{F}$ along the classical trajectories linking the points
$\mathbf{r}^{\prime }$ and $\mathbf{r}$ \cite{Shapoval,dGennes,Luders}. 
(As a matter of fact, there are four trajectories of equal contribution: two direct in time trajectories
for opposite orientation of electron spin plus the two time-reversed trajectories.
The probability of each trajectory is equal to the probability of a definite spin orientation: see \textbf{Appendix\,B} for mathematical details.)
As the dynamics of the spins is purely quantum-mechanical, the ratio 
$\displaystyle \frac{\left\vert \mathbf{r}-\mathbf{r}^{\prime }\right\vert }{v_{F}}$
in the preexponential factor of Eq.(\ref{eq:23}) should be identified with time $t$ in Eq.(\ref{eq:16}): see Eq.(\ref{eq:b4}).

In the opposite, ``dirty", limit the kernel has the following coordinate representation:
  %24=>
\begin{equation}
\mathcal{K}\left( p\right) =2 N\left( 0\right) \left\vert g \right\vert T_{c}\sum\limits_{\omega _{n}>0}%
\frac{2\omega _{n}+\frac{D}{\hbar}p^{2}}{\left( 2\omega _{n}+\frac{D}{\hbar}p^{2}\right) ^{2}+4J^{2}}~, \label{eq:24}
\end{equation}
   %25=>
\begin{eqnarray}
&\mathcal{K}&\left( \mathbf{r,r}^{\prime }\right) =
\frac{N\left( 0\right) \left\vert g \right\vert T_{c}}{\hbar\left\vert \mathbf{r}-\mathbf{r}^{\prime }\right\vert  D} \nonumber \\
&\times& \sum\limits_{\omega _{n}>0}
\left\{ 1 - 2\sin ^{2}\left[ \frac{\left\vert \mathbf{r}-\mathbf{r}^{\prime }\right\vert }{ 2\sqrt{\hbar D}}\sqrt{\sqrt{\omega _{n}^{2}+J^{2}}-\omega _{n}}\right] \right\} \nonumber \\
&\times& \exp \left[  - \frac{\left\vert \mathbf{r}-\mathbf{r}^{\prime }\right\vert}{ \sqrt{\hbar D}} \sqrt{\sqrt{\omega _{n}^{2}+J^{2}}+\omega_{n}}\right]. \label{eq:25}
\end{eqnarray}
Here, $\displaystyle D=\frac{v_{F}l}{3}$ is the diffusion coefficient. As in the ``clean''
limit, the preexponential factor (in the figure brackets) reflects spin-flip
processes. The complexity of the argument of the spin-flip probability 
($\sin ^{2}\left[ ...\right] $)) in Eq.(\ref{eq:25}) is due to the fact that in
the ``dirty'' limit the relevant classical trajectories of electron motion are
those of a random walk process \cite{dGennes,Luders}: see Eq.(\ref{eq:b5}). 

The above equations (\ref{eq:23}) and (\ref{eq:25}) do not exhibit any trace
of the $\Psi _{S}\rightarrow \Psi _{T,S_{z}=0}$ conversions described in the previous
section, because the BCS Hamiltonian precludes the formation of
superconducting correlations between two electrons in a triplet state \cite{Bardeen,Anderson84}.
By contrast, the accompanying effects of the vanishing of the
probability amplitude $a\left( t\right) $ and spin permutations within the
singlet pair do take place. These effects can be interpreted as a
manifestation of a new mechanism of the destruction of superconducting
correlations, completely overlooked in the literature. 
Finally, we want to say a few words about an application of our results to
spin chemistry. 

%SECTION 4.2
\subsection{Spin chemistry}
Spin chemistry \cite{Salikhov,Rodgers,Hore} is a new and rapidly developing interdisciplinary science
relating chemistry, physics and biology. It is concerned with the effect of
external magnetic fields (including static ones) on chemical reactions. A
significant group of chemical reactions, sensitive to external static
magnetic fields, involve as intermediates so-called radical pairs in the
singlet state. Singlet radical pairs themselves emerge, in particular, when
certain organic molecules experience photochemical reactions that are
accompanied by electron transfer from one molecular complex to the other \cite{Rodgers,Hore}.
Although singlet radical pairs are usually short-living and tend to
recombine, it has been noticed that external static magnetic fields can
induce a conversion of the singlet state of radical pairs to the triplet
one. Our exact solution represented by Eqs.(\ref{eq:13})-(\ref{eq:17})
sheds new light on the nature of this latter effect.

Indeed, it is universally believed in spin chemistry \cite{Rodgers} that the
$\Psi _{S}\rightarrow \Psi _{T,S_{z}=0}$ conversion in not too small static
magnetic fields should be ascribed to presumed inequality of spin Land\'{e}
factors of the members of a radical pair (i.e., $\Delta g_{s}\equiv g_{s1}-g_{s2}\neq 0$).
However, Eqs.(\ref{eq:13})-(\ref{eq:17}) suggest that $\Psi _{S}\rightarrow \Psi _{T,S_{z}=0}$
conversions may occur under the condition of equal $g_{s}$-factors for both the members of the radical pair
(i.e., no assumption of the inequality $\Delta g_{s}\neq 0$ is requred). 

Certainly,
the value of the $g_{s}$-factor was calculated by
methods of quantum electrodynamics for free electrons only \cite{Schwinger,Weinberg}.
The unpaired electrons of free radicals are by no means free: different
small interactions within each radical may cause the $g_{s}$-factors to
deviate. Nevertheless, our results must
necessarily be taken into account in any considerations of the effect of
singlet-triplet conversions.

% SECTION 5
\section{Discussion and conclusions}
Summarizing, within the framework of a theoretical model described in the
\textbf{Introduction}, we have studied time evolution of the spin part of the singlet
wave function of two electrons in the presence of external static
homogeneous magnetic and exchange fields. In order to obtain the exact
solution to this quantum-mechanical problem, we have had to revise in
 \textbf{Section 2} the traditional approach \cite{DBohm} to the spin singlet, because it
does not take adequately into account the property of invariance under
rotations of the coordinate system. Basing our own approach in  \textbf{Section 2}
solely on this invariance property and using the theory of spinor invariants
\cite{Cartan,Brinkman}, we have derived the generalized representation of the spin
singlet [Eq.(\ref{eq:2})] whose fundamental feature is that the spins are in
mutually time-reversed states.

We think that exactly the misunderstanding of the above-mentioned
fundamental feature of the spin singlet is the main reason why the problem
of time evolution has not been solved in the available literature. In this
regard, it would be in order to point out that, although the alternative
form of the generalized representation is well-known (at
least, in the theory of superconductivity \cite{LP}), any detailed analysis
of the representation, analogous to ours in \textbf{Appendix\,A}, has not
been undertaken. In particular, the correct form of the interaction
Hamiltonian [our Eqs.(\ref{eq:10}) and (\ref{eq:11})], which is crucial to the
solution of the problem of time evolution, has not been established.

Our exact solution to the problem of time evolution [Eqs.(\ref{eq:13})-(\ref{eq:17})],
derived by different mathematical methods in \textbf{Section 3} and \textbf{Appendix\,A},
implies the existence of two non-trivial quantum-mechanical effects, namely:
periodic singlet-triplet conversions and periodic permutations of the spins within the singlet.
These effects are described in more detail in \textbf{Section 3} itself and \textbf{Section 4}
concerned with some applications to the theory of ferromagnetic
superconductors and spin chemistry.

By the way, the theory of ferromagnetic superconductors provides a very good
illustration of the validity of the exact solution (\ref{eq:13})-(\ref{eq:17}) and
its consequences: the quasi-classical expressions (\ref{eq:23}) and (\ref{eq:25}),
derived by quantum-mechanical methods in \textbf{Appendix\,B}, have as
Fourier transforms the well-known \cite{Saint-James} expressions (\ref{eq:22}) and
(\ref{eq:24}), respectively. However, applications to the theory of
ferromagnetic superconductors by no means reduce to mere restatement of
already known results: one of the implications of the exact solution (\ref{eq:13})-(\ref{eq:17})
is a new mechanism of the destruction of superconducting
correlations by the exchange field, not reported in previous publications.

As regards applications to spin chemistry \cite{Salikhov,Rodgers,Hore}, our exact
solution (\ref{eq:13})-(\ref{eq:17}) yields a natural explanation of the
experimentally observed effect of singlet-triplet conversion in radical
pairs in the presence of external static magnetic fields. This explanation
does not require any assumptions of inequality between the relevant spin Land%
\'{e} factors, which should be contrasted with typical publications on this
subject \cite{Rodgers}: see \textbf{Section 4} for more detail. To draw the line, we
think that our results may stimulate further theoretical studies of the
problem of time evolution of the singlet state of two electrons on the basis
of more realistic models than the one employed in our paper.

\ack
The authors are grateful to A. S. Kovalev for invaluable help in discussing the content of this article.
The authors thank A. I. Korobov for a fruitful discussion of relevant
problems of spin chemistry.

\appendix

\section{SPIN SINGLET AS THE NORMALIZED METRIC SPINOR}

We begin by reminding the well-known \cite{Landau} property of the metric spinor:
   %A1=>
\begin{equation}
g_{ij}=g^{ij}. \label{eq:a1}
\end{equation}%
This property means that the matrix $g$ can be regarded
both as a covariant and a contravariant spinor of rank two, which is verified directly.
Moreover, the metric spinor satisfies a set of elementary relations:
   %A2=>
\begin{equation}
g^{+}=g^{-1}=\widetilde{g}=-g, \label{eq:a2}
\end{equation}%
where the tilde $ \, \, \widetilde{ } \,$ denotes a transposition. 

If we now write down explicit expressions for the direct products of the
spinors on the right-hand side of (\ref{eq:2}) we immediately get:
   %A3=>
\begin{eqnarray}
\Psi _{S} &=&\frac{1}{\sqrt{2}}g,  \nonumber \\
\Psi _{S}\left( i;j \right)  &=&\frac{1}{\sqrt{2}}g^{ij}=\frac{1}{\sqrt{2}}g_{ij}. \label{eq:a3}
\end{eqnarray}%
(This result is just a manifestation of the fact that any antisymmetric
spinor of rank two is equal to the metric spinor multiplied by a scalar \cite{Landau}.)

It is instructive to check the main properties of the spin singlet for
expression (\ref{eq:a3}) independently. The fulfillment of the normalization condition is evident:
   %A4=>
\begin{equation}
\textrm{Sp}\left( \Psi _{S}^{+}\Psi _{S}\right) =-\frac{1}{2}\textrm{Sp}\left(g^{2}\right) =1. \label{eq:a4}
\end{equation}
As is well known from the classical mechanics \cite{Goldstein}, any rotation of the Cartesian
coordinate system about the origin can be parameterized by the Euler angles
and is represented by a product of three consecutive rotations about certain
axes. Therefore, to verify the invariance of (\ref{eq:a3}) under rotations, it
is sufficient to consider rotations by an angle $\phi $ about an arbitrary
axis specified by a unit vector $\left\vert \mathbf{m}\right\vert $ ($%
\left\vert \mathbf{m}\right\vert =1$). The transformation of spinor
components under such rotations are realized by the unitary transformation
matrix \cite{Landau}
   %A5=>
\begin{equation}
D\left( \mathbf{m};\phi \right) =\exp \left( \frac{i}{2}\mathbf{m\vec{\sigma}}\right). \label{eq:a5}
\end{equation}
To avoid misunderstandings, we note that the matrix $D$ is not a spinor; therefore, the position of the matrix indices (upper, lower or mixed) is nonessential for this matrix. Thus, we write:
   %A6=>
\begin{eqnarray}
\Psi _{S}\left( i^{\prime };j^{\prime }\right)
&=&D_{\;k}^{i^{\prime}}D_{\;l}^{j^{\prime }}\Psi _{S}\left( k;l\right) 
\equiv \frac{1}{\sqrt{2}}D_{\;k}^{i^{\prime }}D_{\;l}^{j^{\prime }}g^{kl}\nonumber \\
&=&\frac{1}{\sqrt{2}}D_{\;k}^{i^{\prime }}g^{kl}\widetilde{D}_{l}^{\;\,j^{\prime }}  
=\frac{1}{\sqrt{2}}D_{\;k}^{i^{\prime }}\left[ D^{-1}\right]^{\!k}_{\;l}g^{lj^{\prime }}\nonumber \\
&=&\frac{1}{\sqrt{2}}g^{i^{\prime }j^{\prime }}=\Psi _{S}\left( i;j\right). \label{eq:a6}
\end{eqnarray}
In the above transformations we have used convention concerning the
repeated indices and employed commutation relations
between the Pauli matrices.

A proof of the fact that the spins of a singlet pair are in mutually
time-reversed states is slightly more involved. Consider a somewhat
idealized situation when these spins are separated far apart in the
coordinate space, so that only one of the spins (say, the spin whose state
is specified by the row index of the matrix $g$) is under the influence of
the perturbation, whereas the second one (whose state is
specified by the column index) is not. In this situation, the state of the
pair is described by the time-dependent function
   %A7=>
\begin{eqnarray}
\Psi_{\textnormal{row}} &=& \frac{1}{\sqrt{2}}\left[ U\left( t\right) \right]_{\;k}^{i} g^{kj} \nonumber \\
&=& \frac{1}{\sqrt{2}}g^{ik}\left[ \widetilde{U}\left(-t\right) \right] _{\!k}^{\;j} 
= \frac{1}{\sqrt{2}}\left[ U_{\textnormal{rev}}\left(t\right) \right] _{\;k}^{j} g^{ik} \nonumber \\
&=& -\frac{1}{\sqrt{2}}\left[ U_{\textnormal{rev}}\left(t\right) \right] _{\;k}^{j} g^{ki}, \label{eq:a7}
\end{eqnarray}%
where the evolution operators $U\left( t\right)$ and $U_{\textnormal{rev}}\left(t\right)$
are given by Eq.(\ref{eq:4}) and Eq.(\ref{eq:5}), respectively. In
the last line of (\ref{eq:a7}) the antisymmetry property of the spin singlet has been used.

Similarly, in the opposite situation, when the role of the spins is interchanged, we have:
   %A8=>
\begin{eqnarray}
\Psi _{\textnormal{column}}&\!=\!&\frac{1}{\sqrt{2}}\left[ U\left( t\right) \right]_{\;k}^{j}g^{ik} \nonumber \\
&\!=\!&\frac{1}{\sqrt{2}}\left[ \widetilde{U}\left( -t\right)\right] _{\;k}^{\!i}g^{kj} 
\!=\!\frac{1}{\sqrt{2}}\left[ U_{\textnormal{rev}}\left(t\right) \right] _{\;k}^{i}g^{kj} \nonumber \\
&\!=\!&-\frac{1}{\sqrt{2}}\left[ U_{\textnormal{rev}}\left(t\right) \right] _{\;k}^{i}g^{jk}. \label{eq:a8}
\end{eqnarray}
A comparison between the first and the last lines of relations (\ref{eq:a7}) and (\ref{eq:a8}) proves our time-reversal-symmetry statement.

The above considerations allow us to conclude that in the situation, when both the spins are under the influence of the perturbation, 
their state is represented by either the time-dependent function
   %A9=>
\begin{eqnarray}
\Psi \left( t\right)
&=&\frac{1}{\sqrt{2}}\left[ U\left( t\right) \right]_{\;k}^{i}\left[ U_{\textrm{rev}}\left( t\right) \right] _{\;l}^{j}g^{kl} \nonumber \\
&=&\frac{1}{\sqrt{2}}\left[ U\left( t\right) \right] _{\,k}^{i}g^{kl}\left[\widetilde{U}_{\textrm{rev}}\left( t\right) \right] _{\!l}^{\;j}\nonumber \\
&\equiv& \frac{1}{\sqrt{2}}U\left( t\right) g\widetilde{U}_{\textnormal{rev}}\left( t\right), \label{eq:a9}
\end{eqnarray}
or by its complex conjugate
   %A10=>
\begin{eqnarray}
\Psi^{*} \left( t\right)
&=&\frac{1}{\sqrt{2}}\left[ U_{\textnormal{rev}}\left( t\right) \right]_{\;k}^{i}\left[ U \left( t\right) \right] _{\;l}^{j}g^{kl} \nonumber \\
&=&\frac{1}{\sqrt{2}}\left[ U_{\textnormal{rev}}\left( t\right) \right] _{\;k}^{i}g^{kl}\left[\widetilde{U}\left( t\right) \right] _{\!l}^{\;j}\nonumber \\
&\equiv& \frac{1}{\sqrt{2}}U_{\textnormal{rev}}\left( t\right) g\widetilde{U}\left( t\right). \label{eq:a10}
\end{eqnarray}
Explicitly, these two relations, of course, reproduce relations (\ref{eq:13}) and (\ref{eq:20}) of the main text with $\Psi _{S}$ and $\Psi _{T,S_{z}=0}$ in the matrix form:
   %A11=>
\begin{equation}
\Psi _{S}\equiv \frac{1}{\sqrt{2}}g,\quad \Psi _{T,S_{z}=0}\equiv \frac{1}{\sqrt{2}}\sigma _{z}.  \label{eq:a11}
\end{equation}

The symbolic forms of the last lines of relations (\ref{eq:a9}) and (\ref{eq:a10}) are convenient for the determination of the probability amplitude $a\left( t\right) $:
\begin{eqnarray}
a\left( t\right)
&=&\textnormal{Sp}\left[ \Psi _{S}^{+}\Psi \left( t\right) \right]
=-\frac{1}{2}\textnormal{Sp}\left[ gU\left( t\right) g\widetilde{U}_{\textnormal{rev}}\left( t\right) \right] \nonumber\\
&=&-\frac{1}{2}\textnormal{Sp}\left[ gU\left( t\right) gK^{+}\widetilde{U}\left( -t\right) K\right] 
=\frac{1}{2}\textnormal{Sp}\left[ U\left( t\right) K^{+}U\left( t\right) K\right] \nonumber \\
&=&\frac{1}{2}\textnormal{Sp}\left[ K^{+}\left( t\right) K\left( 0\right) \right]
=\frac{1}{2}\textnormal{Sp}\left[ K\left( -t\right) K^{+}\left( 0\right) \right]\nonumber\\
&=&\frac{1}{2}\textnormal{Sp}\left[ \exp \left( -i\frac{2\sigma _{z}J}{\hbar}t\right) \right]
=\cos \left( \frac{2J}{\hbar} t\right); \label{eq:a12}\\
a\left( t\right) 
&=&\textnormal{Sp}\left[ \Psi _{S}^{+}\Psi ^{\ast }\left(t\right) \right]
=-\frac{1}{2}\textnormal{Sp}\left[ gU_{\textnormal{rev}}\left( t\right) g\widetilde{U}\left( t\right) \right]\nonumber\\
&=&-\frac{1}{2}\textnormal{Sp}\left[ gKU\left( -t\right) K^{+}g\widetilde{U}\left( t\right) \right]
=\frac{1}{2}\textnormal{Sp}\left[KU\left( -t\right) K^{+}U\left( -t\right)\right] \nonumber \\
&=&\frac{1}{2}\textnormal{Sp}\left[K\left( 0\right) K^{+}\left( -t\right)\right]
=\frac{1}{2}\textnormal{Sp}\left[K\left( t\right) K^{+}\left( 0\right)\right]\nonumber\\
&=&\frac{1}{2}\textnormal{Sp}\left[ \exp \left( i\frac{2\sigma _{z}J}{\hbar}t\right) \right]
=\cos \left( \frac{2J}{\hbar}t\right) .  \label{eq:a13}
\end{eqnarray}

\section{THE SUPERCONDUCTING INTEGRAL KERNEL AS A TIME LAPLACE TRANSFORM OF A CLASSICAL
CORRELATION FUNCTION}

The kernel of the integral equation (\ref{eq:21}) in the quasi-classical approximation can be represented in the following form:
   %B1=>
\begin{equation}
\mathcal{K}\!\left( \mathbf{r,r}^{\prime }\right) =\frac{2\pi N\!\left( 0\right) \left\vert g \right\vert
T_{c}}{\hbar}\!\!\sum\limits_{\omega _{n}>0}\int\limits_{0}^{+\infty }\!\!dt\exp \left(
-\frac{2\omega _{n}}{\hbar}t\right) f\!\left( \mathbf{r,r}^{\prime };t\right) .  \label{eq:b1}
\end{equation}%
Here, $f\left( \mathbf{r,r}^{\prime};t\right) $ is a sum of four classical
correlation functions times relevant probability factors and appropriate sign: 
   %B2=>
\begin{eqnarray}
\fl 
f\left( \mathbf{r,r}^{\prime};t\right)
&=\left\langle \delta \left( \mathbf{r}\left( t\right) -\mathbf{r}^{\prime}\right) \delta \left( \mathbf{r}\left( 0\right) -\mathbf{r}\right) \right\rangle
_{p_{1}=p_{F};s_{1z}=\frac{\hbar}{2}\textnormal{sign\,}a\left( t\right) } \frac{\left\vert a \left( t\right)\right\vert}{2} \textnormal{sign\,}a\left( t\right)  \nonumber \\
&+\left\langle \delta \left( \mathbf{r}\left( -t\right) -\mathbf{r}^{\prime}\right) \delta \left( \mathbf{r}\left( 0\right) -\mathbf{r}\right) \right\rangle
_{p_{2}=p_{F};s_{2z}=-\frac{\hbar}{2}\textnormal{sign\,}a\left( t\right) }\frac{\left\vert a \left( -t\right)\right\vert}{2} \textnormal{sign\,}a\left(-t\right) \nonumber \\
&+\left\langle \delta \left( \mathbf{r}\left( t\right) -\mathbf{r}^{\prime}\right) \delta \left( \mathbf{r}\left( 0\right) -\mathbf{r}\right) \right\rangle
_{p_{1}=p_{F};s_{1z}=-\frac{\hbar}{2}\textnormal{sign\,}a\left( t\right) }\frac{\left\vert a \left( t\right)\right\vert}{2} \textnormal{sign\,}a\left( t\right) \nonumber \\
&+\left\langle \delta \left( \mathbf{r}\left( -t\right) -\mathbf{r}^{\prime}\right) \delta \left( \mathbf{r}\left( 0\right) -\mathbf{r}\right)\right\rangle
_{p_{2}=p_{F};s_{2z}=\frac{\hbar}{2}\textnormal{sign\,}a\left( t\right) }\frac{\left\vert a \left( -t\right)\right\vert}{2} \textnormal{sign\,}a\left(-t\right).  \label{eq:b2}
\end{eqnarray}

The four terms on the right-hand side of Eq.(\ref{eq:b2}) represent kinematics of the two electrons of a Cooper pair. Thus, for the time interval
$\displaystyle 0\leq t<\frac{\pi \hbar}{4J}$, the first and the third terms correspond to classical motion of electron
1 from the point $\mathbf{r}$ to the point $\mathbf{r}^{\prime}$, with $\displaystyle\frac{\left\vert a\left( t\right) \right\vert}{2}$ being the probability
 of a definite spin orientation: see the definition of $a\left( t\right) $ in Eqs.(\ref{eq:16}) and (\ref{eq:18}), and the text below Eq.(\ref{eq:19}).
At $\displaystyle t=\frac{\pi \hbar}{4J}$, the right-hand of Eq.(\ref{eq:b2}) goes to zero because of a permutation of spin 1 and spin 2: see the main text. 
This effect should be interpreted as the destruction of superconducting correlations (or Cooper pairs) by the exchange field; hence the reduction of the transition
temperature $T_{c}$ analyzed, e.g., in Refs. \cite{Saint-James} and \cite{Izyumov}. As a result of the spin permutation, the function 
$f\left( \mathbf{r,r}^{\prime};t\right)$ acquires minus sign in the time interval $\displaystyle \frac{\pi \hbar}{4J}<t<\frac{3\pi \hbar}{4J}$.
This process is periodic with the period $\displaystyle T=\frac{\pi \hbar}{J}$.

Given time-reversal symmetry of classical mechanics \cite{Messiah} and the equality $a\left( -t\right) =a\left( t\right)$,
it is clear that all the four terms on the right-hand side of Eq.(\ref{eq:b2}) yield equal contributions.
Therefore, Eq.(\ref{eq:b2}) can be rewritten in a more economical form:
   %B3=>
\begin{equation}
f\!\left( \mathbf{r,r}^{\prime };t\right) \!=
\!2\left\langle \delta \!\left( \mathbf{r}\left( t\right) -\mathbf{r}^{\prime }\right) \delta 
\!\left( \mathbf{r}\left( 0\right) -\mathbf{r}\right) \right\rangle _{p=p_{F}}
\!\left\langle K\!\left( t\right) \!K^{+}\!\left( 0\right) \right\rangle \!,  \label{eq:b3}
\end{equation}
where
\begin{equation*}
\left\langle K\left( t\right) K^{+}\left( 0\right) \right\rangle 
\equiv \frac{1}{2}\textnormal{Sp}\left[ K^{+}\left( t\right) K\left( 0\right) \right]
=a\left( t\right) .
\end{equation*}%
[By rewriting the probability amplitude $a$ in the form of a correlator $\left\langle KK^{+}\right\rangle$
we just want to remind that correlators of this kind were first introduced in de Gennes' formulation of the 
theory of superconductivity \cite{dGennes} to describe the effect of non-time-reversal perturbations of different types.
In our case, this correlator is responsible for the preexponential factors in Eqs.(\ref{eq:23}) and (\ref{eq:25}).]

The classical correlator $\left\langle \delta \delta \right\rangle$ is well-known \cite{dGennes,Luders} for the two limiting cases discussed in our paper.
Thus, in the "clean" limit, it reads:%
   %B4=>
\begin{eqnarray}
\left\langle \delta \left( \mathbf{r}\left( t\right) -\mathbf{r}^{\prime}\right) \delta \left( \mathbf{r}\left( 0\right) -\mathbf{r}\right)\right\rangle _{p=p_{F}} \quad ~  \nonumber \\
=\frac{1}{4\pi }\left\vert \mathbf{r}-\mathbf{r}^{\prime }\right\vert^{-2}\delta \left( \left\vert \mathbf{r}-\mathbf{r}^{\prime }\right\vert -v_{F}t\right) .  \label{eq:b4}
\end{eqnarray}%
In the "dirty" limit,
\begin{eqnarray}
\left\langle \delta \left( \mathbf{r}\left( t\right) -\mathbf{r}^{\prime}\right) \delta \left( \mathbf{r}\left( 0\right) -\mathbf{r}\right)\right\rangle _{p=p_{F}} \quad ~   \nonumber \\
=\left( 4\pi Dt\right)^{-\frac{3}{2}} \exp \left[ -\frac{\left\vert \mathbf{r}-\mathbf{r}^{\prime }\right\vert ^{2}}{4Dt}\right].  \label{eq:b5}
\end{eqnarray}%
Upon the substitution of relations (\ref{eq:b4}) and (\ref{eq:b5}) into (\ref{eq:b3})
and carrying out integration over time, we arrive at relations
(\ref{eq:23}) and (\ref{eq:25}) of the main text.

\texttt{ }

\noindent\texttt{---------------------------}

\end{document}